%% file: author.tex
\documentclass[graybox]{svmult}
\usepackage{mathptmx}       % selects Times Roman as basic font
\usepackage{helvet}         % selects Helvetica as sans-serif font
\usepackage{courier}        % selects Courier as typewriter font
\usepackage{type1cm}        % activate if the above 3 fonts are
\usepackage{makeidx}         % allows index generation
\usepackage{graphicx}        % standard LaTeX graphics tool
\usepackage{multicol}        % used for the two-column index
\usepackage[bottom]{footmisc}% places footnotes at page bottom
\usepackage{newtxtext,newtxmath}

\makeindex             % used for the subject index
\begin{document}
\setcounter{chapter}{7}
\title{Towards Synergistic Teacher-AI Interactions with Generative Artificial Intelligence}
\author{
Mutlu Cukurova\inst{1}\thanks{Corresponding author: \email{m.cukurova@ucl.ac.uk}}\orcidID{0000-0001-5843-4854} \and \\
Wannapon Suraworachet \inst{1}\orcidID{0000-0003-3349-4185}\and \\
Qi Zhou \inst{1}\orcidID{0000-0002-4694-4598} \and \\
Sahan Bulathwela\inst{2}\orcidID{0000-0002-5878-2143}}
\institute{
$^{1}$ UCL Knowledge Lab, Institute of Education, University College London, UK, 
\email{\{m.cukurova, wannapon.suraworachet.20,
qtnvqz3\}@ucl.ac.uk}
\and \\$^{2}$ UCL Centre for Artificial Intelligence, Department of Computer Science, University College London, UK,
\email{m.bulathwela@ucl.ac.uk}
}

\maketitle

\abstract*{Generative artificial intelligence (GenAI) is increasingly used in education, posing significant challenges for teachers adapting to these changes. GenAI offers unprecedented opportunities for accessibility, scalability and productivity in educational tasks. However, the automation of teaching tasks through GenAI raises concerns about reduced teacher agency, potential cognitive atrophy, and the broader deprofessionalisation of teaching. Drawing findings from prior literature on AI in Education, and refining through a recent systematic literature review, this chapter presents a conceptualisation of five levels of teacher-AI teaming: transactional, situational, operational, praxical and synergistic teaming. The framework aims to capture the nuanced dynamics of teacher-AI interactions, particularly with GenAI, that may lead to the replacement, complementarity, or augmentation of teachers’ competences and professional practice. GenAI technological affordances required in supporting teaming, along with empirical studies, are discussed. Drawing on empirical observations, we outline a future vision that moves beyond individual teacher agency toward collaborative decision-making between teachers and AI, in which both agents engage in negotiation, constructive challenge, and co-reasoning that enhance each other’s capabilities and enable outcomes neither could realise independently. Further discussion of socio-technical factors beyond teacher-AI teaming is also included to streamline the synergy of teachers and AI in education ethically and practically.
\keywords{Generative AI $\cdot$ Synergy $\cdot$ Teacher-AI Interaction $\cdot$ Teacher Agency}}

\abstract{Generative artificial intelligence (GenAI) is increasingly used in education, posing significant challenges for teachers adapting to these changes. GenAI offers unprecedented opportunities for accessibility, scalability and productivity in educational tasks. However, the automation of teaching tasks through GenAI raises concerns about reduced teacher agency, potential cognitive atrophy, and the broader deprofessionalisation of teaching. Drawing findings from prior literature on AI in Education, and refining through a recent systematic literature review, this chapter presents a conceptualisation of five levels of teacher-AI teaming: transactional, situational, operational, praxical and synergistic teaming. The framework aims to capture the nuanced dynamics of teacher-AI interactions, particularly with GenAI, that may lead to the replacement, complementarity, or augmentation of teachers’ competences and professional practice. GenAI technological affordances required in supporting teaming, along with empirical studies, are discussed. Drawing on empirical observations, we outline a future vision that moves beyond individual teacher agency toward collaborative decision-making between teachers and AI, in which both agents engage in negotiation, constructive challenge, and co-reasoning that enhance each other’s capabilities and enable outcomes neither could realise independently. Further discussion of socio-technical factors beyond teacher-AI teaming is also included to streamline the synergy of teachers and AI in education ethically and practically.
\keywords{Generative AI $\cdot$ Synergy $\cdot$ Teacher-AI Interaction $\cdot$ Teacher Agency}}

\section{Introduction}
\label{sec:1}
Generative Artificial Intelligence (GenAI) is a type of Artificial Intelligence that allows the creation/generation of content based on large volumes of data from various sources. While Large Language Models (LLMs) such as ChatGPT, Microsoft Co-pilot and Google Gemini, that support a text-based conversational interface are primarily referred to as GenAI, GenAI models can also be multi-modal, with capabilities to generate other non-textual modalities such as audio, video and simulations \cite{uk2025generative}.

GenAI has been integrated into human daily life, recording one of the fastest adoption rates of one million users in 5 days \cite{dwivedi_opinion_2023} while holding approximately 800 million weekly users as of late 2025. This trend surpasses any other technological platform of this kind that has ever been developed. With this level of transformation at scale, GenAI has exponentially been utilised in educational settings, especially supporting teachers in various tasks, generating lesson plans \cite{celik_co-constructing_2026}, generating context-specific quizzes \cite{li2025novel} and media \cite{wang_generative_2023, li2024adultlearnersrecallrecognition}, adaptively tutoring students \cite{stienstra_exploring_2025}, scoring assignments \cite{pecuchova_automated_2025} and offering target and specific feedback to students \cite{dai_assessing_2024}. While the integration of GenAI to perform these educational tasks is beneficial in ways that enhance the quality of teaching \cite{dai_assessing_2024} and decrease teachers’ preparation time \cite{roy_chatgpt_2024}, there are habitual concerns pertaining to fully offloading tasks to GenAI, including diminishing teachers’ cognitive improvement \cite{felix_use_2024} and dehumanising learning (as in \cite{edwards_why_2018}) stemming from a lack of communication between teachers and students. 

These benefits and concerns, particularly on teachers’ competence, are influenced by how GenAI is meaningfully integrated in educational contexts. As dominant GenAI models are commercialised primarily with business intent, with no particular design inspiration for education, teachers may have limited control over how this technology can be pedagogically integrated into practice while diminishing teacher agency. \emph{Teacher agency} refers to proactive behaviours and intentions of teachers in actively engaging in decision-making and taking control of their own professional practice, which could positively affect learners’ development as well as contribute to educational contexts \cite{Biesta18082015}. When GenAI is being introduced into the classroom, teachers are expected to have the agency and motivation to decide and adapt pedagogical interventions and educational environments to suit students’ needs as well as their professional goals. This stage of productively integrating human and AI in a context to achieve higher performance than either human or AI could achieve independently is referred to as \emph{Synergy}\cite{vaccaro_when_2024}. While research has begun to conceptualise and identify how humans can positively engage with AI, positioning it as a learning companion \cite{perez2021ai} to collaborate, fostering hybrid intelligence combining human and AI intelligence, there remains a limited understanding of the conditions under which teacher and AI interactions move beyond task operation towards achieving true synergy. 

Tackling from the interaction paradigm, this chapter thereby offers critical perspectives in conceptualising how teachers and AI interact in operationalising certain educational tasks. We establish five different levels of interactions, that is, teacher-AI teaming, that are likely to fuel or prohibit teachers’ professional competency. Then, we further analyse the presence of teacher-AI teaming in existing AI in Education studies from the literature to explore the validity of the proposed framework. Analysing these interactional perspectives can reveal the design affordances and constraints of GenAI tools that shape teacher-AI complementarity through varying levels of teacher agency. Beyond the technological design of what GenAI offers, we also discuss other socio-technical factors that contribute to the effectiveness of teacher-AI interactions. Ultimately, these insights could inform design guidelines for pedagogical GenAI that support teacher agency, while also clarifying how professional development can prepare teachers to engage with GenAI competently, ethically and sustainably.

\section{Generative AI as an Agent}
\label{sec:2}
AI has been significantly gaining attraction in education. The methods proposed and exploited in education have ranged from traditional AI (e.g., search and planning algorithms), classical machine learning (e.g., item response theory, Bayesian Knowledge Tracing) and deep learning (from deep knowledge tracing to LLMs). These approaches predominantly operate on model building and inference to make sense of information, yielding insights that contribute to understanding learning processes and supporting education. A new era of AI, GenAI, has offered a leap in transforming AI capabilities from a detection-/analysis-focused stance to a generation-focused one. GenAI models such as LLMs and Large Multimodal Models (LMMs) are trained on vast datasets, enabling them to understand the underlying patterns and relationships to ultimately produce novel, coherent outputs that mimic human artefacts, including texts, images and other formats \cite{yan_practical_2023}. Moving beyond the predictive and analytic functions of classical AI, GenAI introduces opportunities for co-creation while acting as a reflective companion in teaching practice. With degrees of human language understanding, language-based GenAI systems can be interacted with through initiative, human-like prompts or inputs beyond existing programming languages. The humanly intuitive grounding of communication and instruction following capabilities of GenAI makes it massively accessible to society in contrast to previous waves of AI. This affords more inclusive and accessible opportunities for teachers, especially considering the overwhelming majority of teachers are non-technical users trying to operate and tailor external tools to perform educational tasks that align with their pedagogical goals.

With GenAI’s ability to understand semantic context and predominantly communicate in human language, its behaviours increasingly resemble those of an intelligent agent capable of autonomous interaction. An agent is a system that perceives its environment and autonomously acts upon it to complete/achieve its goals \cite{russell_artificial_2016}. \cite{russell_artificial_2016}, a seminal work in understanding agentic behaviours of intelligent systems, propose five types of agents. This is a well-adopted framework in computer science and educational contexts to categorise different types of AI systems according to their advanced capabilities. These five agent types are: simple-reflex, model-based, goal-based, utility-based and learning agents. In brief, a simple-reflex agent relies on explicit rules written in conditional instructions to be executed, and a model-based agent holds the capabilities to recognise internal states of a situated environment (i.e. model) and act accordingly based on the model. A goal-based agent develops the capability to optimise their actions according to goals (with a long-term destination), whereas a utility-based agent offers higher capabilities in micromanaging a utility function that can allow quantifying the reward for taking certain action sequences towards a goal, such as a certain set of criteria to optimise its actions. At last, a learning agent shows superior learning capability through experience by evolving itself over time, yielding improved performance.

GenAI offers sufficient technical capabilities covering various agent types. While exceeding the capabilities of simple reflex agents, GenAI models could be mainly considered as model-based agents with the ability to maintain prior contextual understanding of users. They also align with goal-based agents in showing capabilities in executing actions according to users’ prompts (goals that can be longer term than what is addressed in the immediate response/utterance). For example, when a GenAI chatbot is being instructed by teachers to summarise students’ discussions, it can take turns to refine the need/goal of the teacher to ultimately deliver a useful output. Alternatively, GenAI models partially fit utility-based agents as they can orient their outputs towards subtle indicators, such as modifying a generated lesson plan to specifically fit a given context, yet it does not achieve this autonomously. Similarly, GenAI models can also become a learning agent as they can learn from a newly trained dataset to further improve performance through a fine-tuning approach; however, this process requires deliberate fine-tuning, not currently autonomously achievable. Indeed, there are recent studies that propose that LLMs use the attention function to do soft learning, which can be considered a form of training/fine-tuning \cite{NEURIPS2023_73950f0e}. Therefore, considering the complexity and diversity in GenAI models, without specifics of a given GenAI model and its operationalised system details in a given task, it is difficult to make a categorisation. In addition, perhaps more importantly for the scope of this chapter, while this agent framework offers a fundamental conceptualisation of AI models, contributing to the understanding of AI models' technical affordances and constraints, there is a lack of explanation on how multiple agents, teachers and AI models interact in systems, and more specifically, those used in educational contexts.

\section{From Single-Agent Characterisation to Human-AI Hybrid Systems}
\label{sec:3}

Extending the theory of distributed cognition \cite{hollan_distributed_2000}, in which cognitive processes are shared across multiple agents, \emph{hybrid intelligence} can potentially emerge when human and AI agents engage in synergy. This is achieved by joining their cognitive capabilities to achieve outcomes that neither one of them could accomplish individually at the level of their performance as a team. Several conceptual frameworks have been proposed on how AI and teachers cooperate in educational settings to achieve hybrid intelligence. For instance, \cite{bittencourt_conceptual_2020} conceptualises a hybrid complementarity framework, considering four aspects of teacher competence that are supported by AI, namely goal setting, perception, actions and decisions. In other words, AI could help support teachers in (1) setting and evaluating their pedagogical goals, (2) enhancing their perception of classroom situations and students’ learning processes, (3) scaling pedagogical task delivery or operations, and (4) offering a decision support system or recommendations based on data.
Although this framework offers a useful starting point to determine target areas of teacher enhancement supported by AI, it considers the teacher-AI combined space as static, with insufficient information on their dynamic interactions. 

Conversely, \cite{molenaar_towards_2022} considers teacher-AI interactions from the aspect of the level of automation and control between teachers and AI, adapting the six levels of automation from a broader industrial and medical context to educational contexts. While this sheds light on the human-AI interactions with varying degrees of control in achieving educational tasks, such a mapping leaves limited room for understanding how AI helps develop teacher competency from a cognitive perspective and how AI learns from the interactions to further evolve. This is because the framework is strictly oriented towards task division between the teacher and AI agents, promoting productivity, as commonly prioritised in other domains, but may not necessarily align with educational goals, which are oriented towards cognitive improvement.  

Cukurova \cite{cukurova_interplay_2025} posits three levels of AI integration along the spectrum of AI automation vs human control and their impacts on teacher competency in educational contexts: replacement, complementarity, and augmentation. \emph{Replacement} refers to a situation where AI systems independently operate teaching tasks without, or with minimal, teacher oversight (high AI automation with low teacher control). Although replacement can lead to better performance at task completion, it is likely to lead human competence to flatten or decline. \emph{Complementarity} offers an alternative conceptualisation of AI in education, in which AI systems enhance teachers’ capabilities through supportive information or operations with active control from teachers (low AI automation with high teacher control). The goal of a complementary system is not only to help improve performance, but also help humans maintain or improve their own competence. \emph{Augmentation}, on the other hand, offers a conceptualisation in which greater competence of teachers and AI is achieved through interwoven interactions between teachers and AI. In augmentation the goal is to achieve performance that is superior than the maximum of each agent alone and improve the competence of both agents in the process (high AI automation with high teacher control). Although this conceptualisation offers a concrete mapping of potential teacher-AI interactions and their implications for teacher competence, it does not clarify how such interactions unfold at a granular level, offering limited guidance on how teacher-AI interactions should be productively designed. Collectively, there is a need to establish a conceptual framework to systematically analyse types of interactions between both agents to investigate cognitive impacts of both agents towards co-adaptation, which could potentially shape the synergistic paradigm of teacher competence development through their interactions with AI.

\section{Five Levels of Teacher-AI Teaming in Educational Contexts}
Considering interaction affordances between teachers and AI, supporting their cognitive enhancement and competency development, we propose an orthogonal conceptual framework (contrary to task-oriented interpretation of automation) on teacher-AI teaming, focusing on information that flows, decisions and actions that are operated and shared between the two agents (Human and AI) in operating teaching tasks and supporting cognitive capabilities of both agents. 

The framework proposed here originates from \cite{crowley_hierarchical_2023}' human-AI collaboration framework. Considering from the viewpoint of agentic cognitive capabilities, \cite{crowley_hierarchical_2023} proposes a hierarchical framework: reactive, situational, operational, praxical and creative collaboration, offering a broad concept of human and AI interactions along the dependency and complexity of \emph{collaborative AI} systems. While this framework offers a promising starting point, due to its alignment with \cite{russell_artificial_2016}'s characterisation of intelligent agents at the cognitive level, it argues for human-AI collaboration. This reflects its philosophical stance of viewing AI as a comparable agent that can indeed \emph{collaborate} with humans. The framework also lacks a strict underpinning for the educational context but instead gains inspiration from engineering contexts with an operational, output-oriented focus. Educational contexts, on the other hand, tend to focus on the cognitive improvement of agents rather than a mere focus on task completion and efficiency. We thereby adopted \cite{crowley_hierarchical_2023}'s framework to formulate our teacher-AI interaction conceptualisation grounded in educational contexts. We choose to use the term \emph{teaming} rather than \emph{collaboration}, as collaboration is a well-established concept with clear definitions and design features of interactions (e.g. mutuality, interdependency, equality etc.) in research communities like computer supported collaborative learning \cite{dillenbourg_what_1999} and peer education \cite{damonCriticalDistinctionsThree1989}. It is not clear to what extent human-AI interactions can be categorised as collaboration. In this chapter, we refine our framework iteratively by validating the presence of the different teaming levels identified in the literature on AI supports for teachers.

\begin{figure}[ht]
    \centering
    \includegraphics[width=1\textwidth]{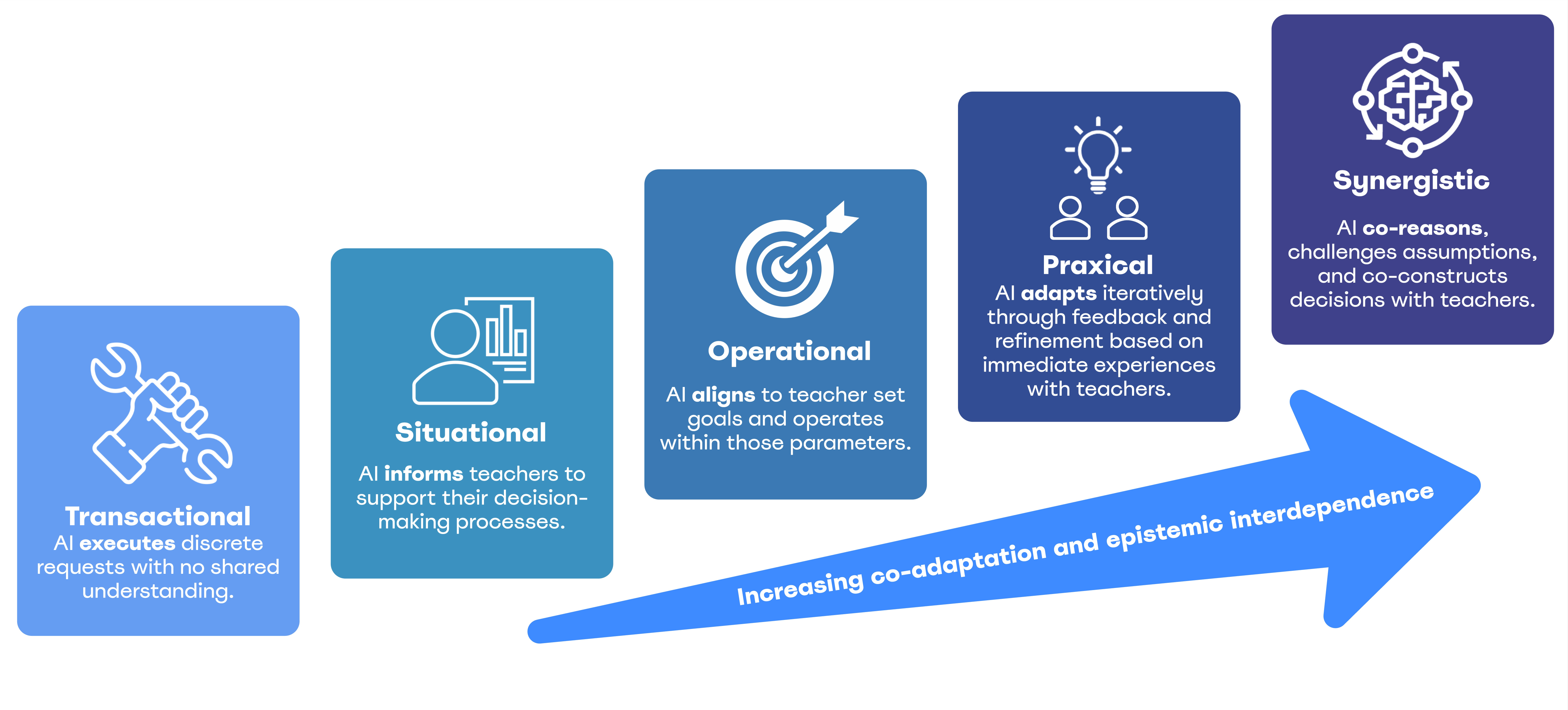}
    \caption{The five levels of teacher-AI teaming}
    \label{fig:fiveTeaming}
\end{figure}

The teaming levels start from minimal (transactional) teacher-AI interactions towards synergistic interactions between teachers and AI. Specifically, we propose (1) transactional, (2) situational, (3) operational, (4) praxical and (5) synergistic levels of teaming,  as shown in Fig. \ref{fig:fiveTeaming}. In the subsequent sections, we describe these levels in detail, supported by a teacher support application-based example.

\begin{figure}[ht]
    \centering
    \includegraphics[width=1\textwidth]{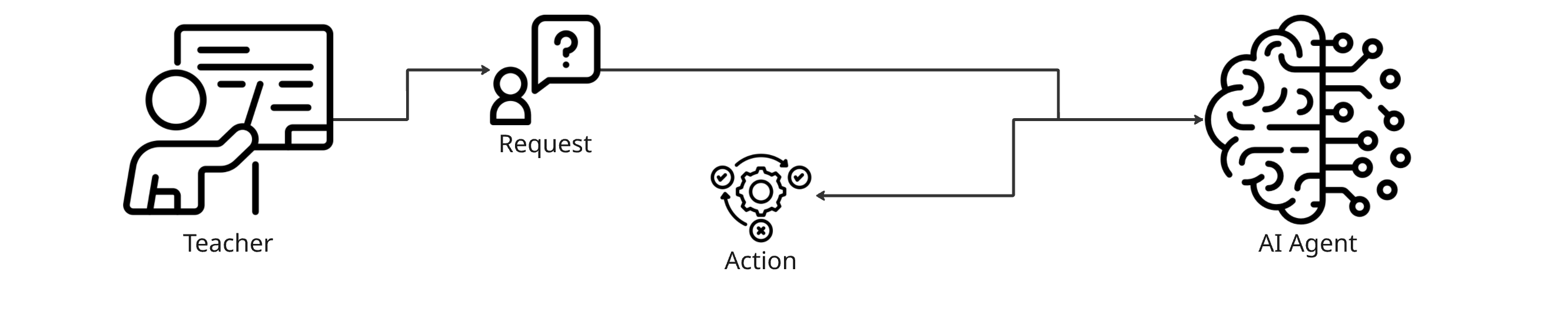}
    \caption{The transactional teaming. A teacher sends a request to an AI agent, and the AI agent then performs the requested action(s). Adapted from \cite{oecdOECDDigitalEducation2026}.}
    \label{fig:transactional}
\end{figure}

\textbf{Transactional level} entails a straightforward request-response mechanism being deployed within educational contexts where teachers instruct AI to perform a certain task autonomously following teachers’ discrete requests, as shown in Fig. \ref{fig:transactional}. In other words, a teacher provides an input command/instruction, and AI then executes and outputs a result. Transactional teaming usually happens with repetitive, minuscule, non-pedagogically focused tasks where the task is sufficiently concrete to be automated. AI in a transactional team serves as a complete outsourcer for teachers to improve productivity, rather than enhancing the cognitive capabilities of either agents.

To illustrate, a teacher requests AI to translate an input text to another language for students could be perceived as a transactional teaming scenario, as the outcome of the same instruction varies very little based on background variables.

\begin{figure}[ht]
    \centering
    \includegraphics[width=1\textwidth]{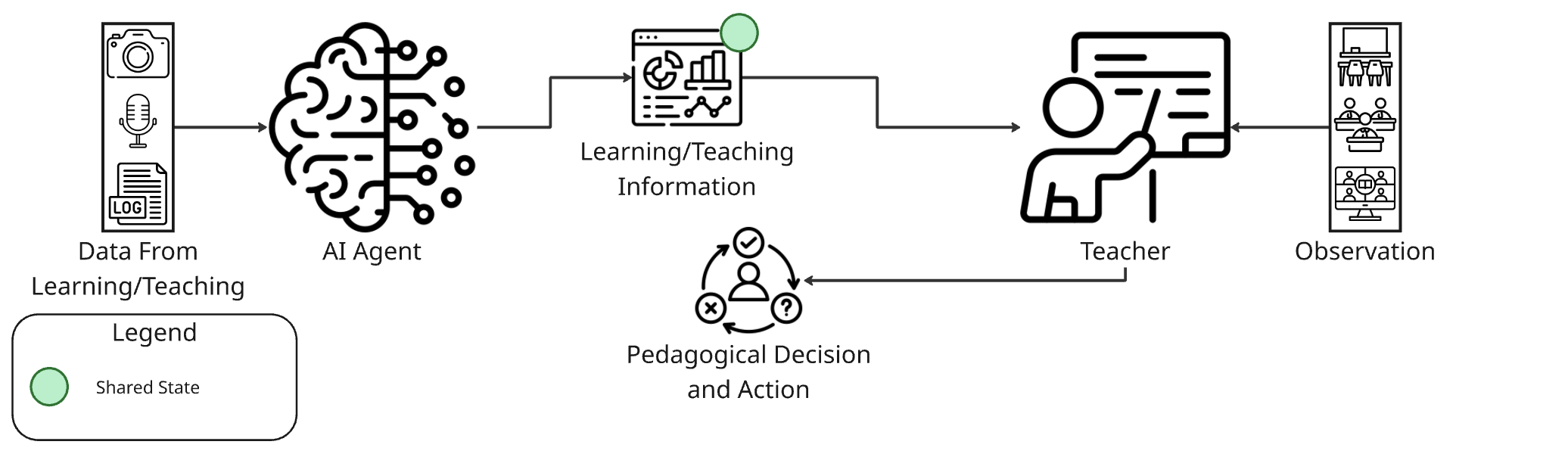}
    \caption{The situational teaming. An AI agent gathers data from learning/teaching contexts. It then sends information to teachers. A teacher then combines their own observation with AI information to perform pedagogical decisions/actions(s). Adapted from \cite{oecdOECDDigitalEducation2026}.}
    \label{fig:situational}
\end{figure}

\textbf{Situational teaming} refers to interactions between teachers and AI to support shared awareness of the teaching and learning context, as shown in Fig. \ref{fig:situational}. The key differentiator of situational teaming from transactional teaming is that there is a shared state for the task that changes the agentic behaviours (may it be the human or AI agent). In essence, AI and teachers communicate to offer this state information to one another. Through this complementary information, teachers could potentially enhance their cognition.

For instance, AI systems gather contextual data from classrooms and/or learning activities retrieved from online learning systems, physical sensors or data streams, interpreting using internal models and offering educationally meaningful information to teachers. Based on AI-generated insights, teachers thereby take full agency in integrating information to make informed decisions and deliver suitable pedagogical interventions accordingly.

\begin{figure}[hbt!] 
    \centering
    \includegraphics[width=1\textwidth]{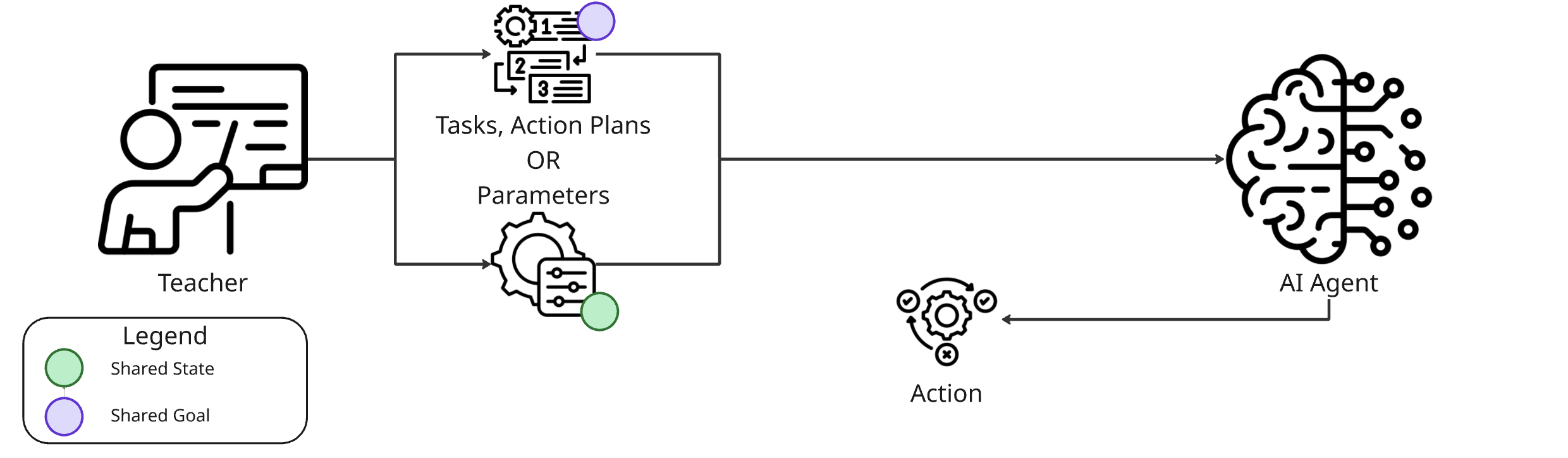}
    \caption{The operational teaming. A teacher sets a goal(s) for an AI agent in terms of tasks, action plans or parameters, and the AI agent then performs the requested action(s) according to the teacher's instructions. Adapted from \cite{oecdOECDDigitalEducation2026}.}
    \label{fig:operational}
\end{figure}

\textbf{Operational teaming} targets task division between teachers and AI. This includes teacher-AI cooperation of plans and teaching operations deployed in teaching/learning contexts, as shown in Fig. \ref{fig:operational}. The \emph{plan} is essential to entail a shared goal that is long-term, going beyond the current context, which distinguishes operational teaming from situational teaming. 

To illustrate, the interaction starts when teachers provide instructions or parameters for AI systems about their goals or assigned tasks. The AI system then integrates these goals into its decision-making and adaptively acts following teachers’ requests in executing autonomous tasks or supporting teachers with goal-oriented information. This teaming extends beyond a mere task execution (happening at the transactional teaming) or reporting of situations (happening at the situational teaming) for teachers towards higher teacher control in assigning pedagogy-related goals for AI systems, aligning with teachers’ needs and enabling pedagogically-grounded task execution. Most current prompt engineering interactions of teacher in genAI use can potentially be categorised in this operational teacher-AI teaming level.

\begin{figure}[ht]
    \centering
    \includegraphics[width=1\textwidth]{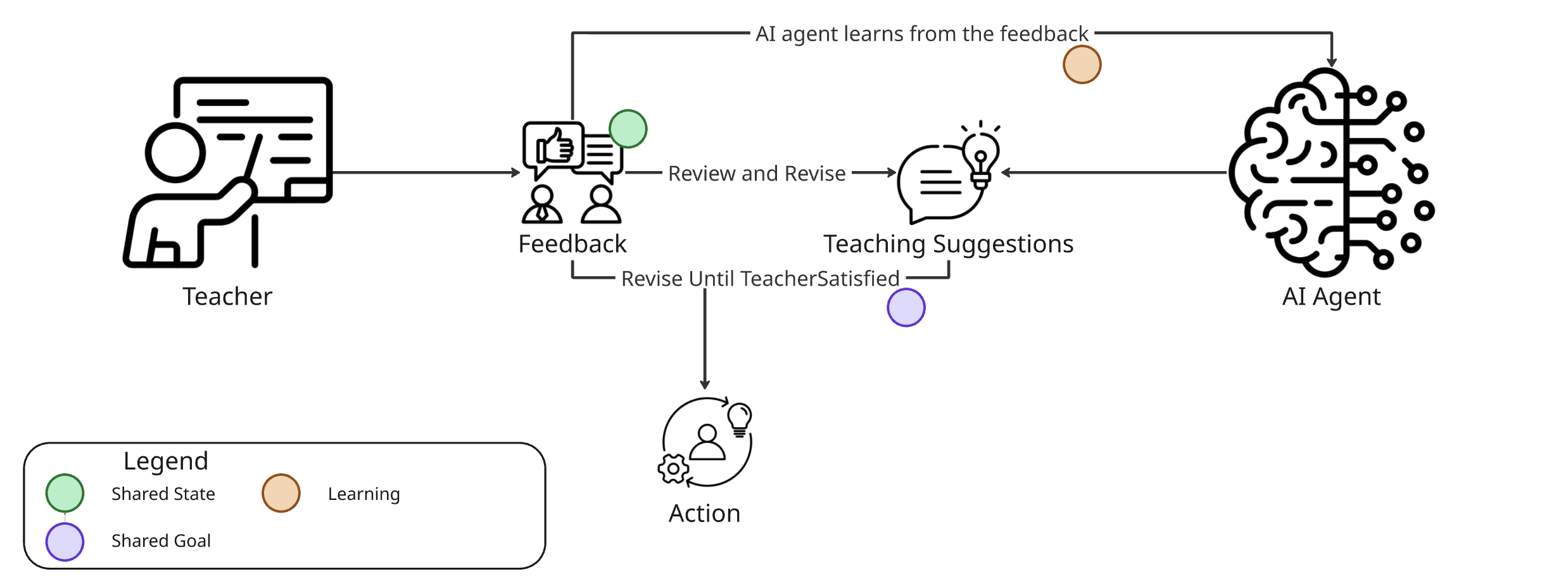}
    \caption{The praxical teaming. An AI agent provides feedback/information about learning/teaching contexts to a teacher. The teacher reviews and revises the feedback. The AI agent then incorporates the teacher's inputs to produce new feedback. The teacher and the AI agent can engage in this iterative reviewing and revising process until the teacher is satisfied with the feedback before they take actions. Adapted from \cite{oecdOECDDigitalEducation2026}.}
    \label{fig:praxical}
\end{figure}

\textbf{Praxical teaming} emphasises that teachers and AI systems dynamically exchange information on teaching and learning procedures, actions or plans towards establishing a shared understanding and effective practice, as shown in Fig. \ref{fig:praxical}. This means that the agents can learn from each other to change their behaviours to optimise the collective system over time. 

While teachers activate their prior experience and professional expertise through externalising their skills and knowledge in their edits/feedback to AI during praxical teaming, AI systems incorporate learning patterns or information queried from databases to tackle specific tasks/questions at hand. Essentially, AI systems can learn from teacher feedback (implicit, such as choices, preferences and explicit, such as liking, rating) by updating their model and adapting to teachers’ pedagogical guidance or preferences in the long term.

\begin{figure}[ht]
    \centering
    \includegraphics[width=1\textwidth]{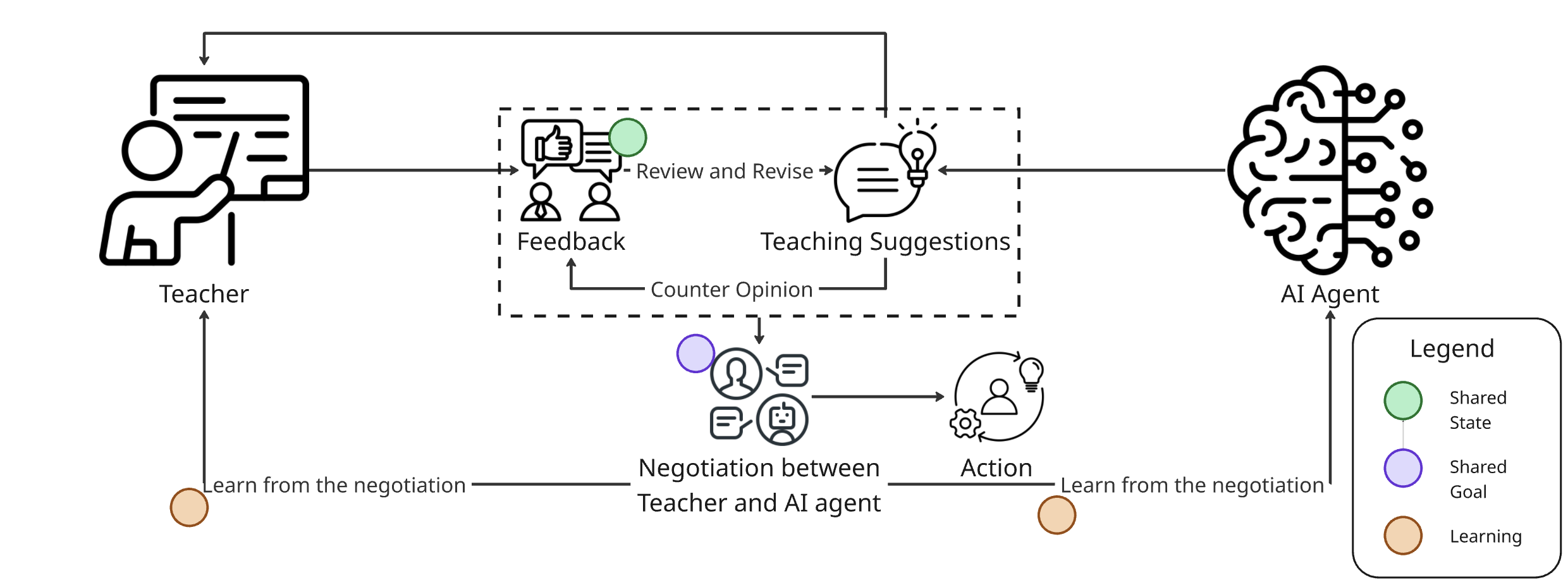}
    \caption{The synergistic teaming. Similar to the praxical level, an AI agent provides feedback/information about learning/teaching contexts to a teacher. The teacher reviews and revises the feedback. Rather than the AI agent obediently incorporating the teacher's inputs to revise feedback, the AI agent will critically engage in challenging the ideas of teachers by providing counter opinions. Both agents iteratively negotiate until reaching a consensus. Adapted from \cite{oecdOECDDigitalEducation2026}.}
    \label{fig:synergistic}
\end{figure}

\textbf{Synergistic} teaming happens when teachers and an AI system maintain their agency while mutually engaged in co-adaptation, as shown in Fig. \ref{fig:synergistic}. In contrast to praxical teaming, the AI system operated at a synergistic teaming level should not only blindly follow teachers’ instructions (operational teaming) and adapt to teacher feedback (praxical teaming), but also be able to critically engage in evaluating teacher feedback, reason with logic and evidence, and where necessary, \emph{push back} or offer a gentle nudge when teacher instruction is incongruent with the AI agent's self-evaluation of the best course of action. This teaming requires high levels of agency from both teachers and AI systems to critically debate and extend the boundaries of one another, reflecting dynamic and prolonged communication and co-reasoned decisions among agents (such as through representation sharing, uncertainty negotiation, initiative management, explanation-for-action etc.). When this level of teaming is achieved, it forms a creative resonance and enables teachers and AI systems to deepen task understanding and potentially generate innovative outcomes that individual agents could not achieve alone \cite{cukurova_interplay_2025}, signifying the potential synergistic augmentation of teacher competence towards hybrid intelligence. In comparison to the synergistic teaming, the praxical teaming relies on teachers’ current competence to give feedback to AI. With a high reliance on teacher feedback, the combined competence of both agents often saturates at the maximum competence of teachers, leaving little or no improvement beyond their current practices. Not limited by teacher oversight, synergistic teaming supports co-reasoned decisions where both parties should interact to explain reasoning and share decisions towards the direction of the collective outputs, ultimately resulting in co-adaptation of both agents.

While the proposed five levels of teaming represent some ordered characteristics, these need not always follow a progressive ordered hierarchy (e.g., one can indeed learn from others' behaviours without having a shared goal). Similarly, improving the teaming level does not guarantee improving teaching performance, but rather it represents higher-level cognitive exchanges among the two parties, as well as higher potential for improving their cognitive abilities towards co-adaptation. Grounded in distributed cognition \cite{hollan_distributed_2000} and extended cognition \cite{clark_1998} theories, teacher competency could potentially be complemented and augmented maximally at the praxical and synergistic levels. However, educational contexts are fuzzy and complex, which require context-sensitive teaming to be appropriately operated upon in a given task rather than aiming for synergy in every human-AI interaction. 

\section{The Presence of Teacher-AI Teaming Levels in GenAI Applications}
Having established a framework for human-AI teaming, it is necessary to examine how effectively existing teacher support tools can be interpreted through this lens. To examine the framework’s applicability, we conducted a systematic review of empirical studies on AI tools supporting teachers in educational contexts (teacher-facing AI tools) from 2010 to 2025, following the guidelines of \cite{kitchenham_guidelines_2007}. An initial keyword search of the Web of Science database, combining AI-related and teacher-related terms within educational contexts, retrieved over 7000 articles, which were independently screened by title and abstract by two coders, achieving substantial agreement (Cohen’s K = 0.82). Four coders then performed data extraction following a pilot phase to ensure consistency, achieving moderate agreement (Krippendorff’s alpha = 0.69). Any disagreements were resolved through discussion. Full extraction resulted in 103 studies. Information on AI type and teacher-AI teaming levels was recorded, enabling a secondary analysis of GenAI-powered teacher-facing tools, representing 39\% of the included studies. The following paragraphs describe examples of the empirical studies operating on GenAI in the teaching context. To illustrate the framework more clearly, we confine our examples to text-based conversational systems, such as chatbots.

At the \textbf{transactional level} (17.5\%), teachers interact with GenAI systems through direct prompts, typically via chatbots, receiving generated responses as outputs. For instance, in mechatronic virtual laboratories, teachers can prompt a customised GPT model to analyse preprocessed numerical sequences from experimental steps; the resulting outputs are then compared with students’ patterns to support an automatic correction system for both teachers and students \cite{machado_automatic_2025}. Transactional interaction in which teachers trigger or request AI likely dominates current interactions \cite{handa_which_2025}, partly due to its benefits in automating ranges of generation tasks in educational practice, which were unachievable before GenAI due to the task-specific nature of conventional AI. These transactional interactions are less prominent in our review, primarily because we focused on studies examining teacher-AI interactions rather than those in which AI autonomously performs complete teaching tasks. Orienting towards task automation, teacher actions exhibited in this teaming include initiating a request and verifying GenAI responses, whereas exchanges at the cognitive level remain limited as productivity and efficiency are prioritised \cite{zhang_evaluating_2025}.

GenAI leveraged at the \textbf{situational teaming} (17.5\%) orients towards offering situational awareness to teachers in capturing data streams from learners and/or learning environments regarding learning activities and learning processes, analysing and providing meaningful insights to teachers. Situational teaming targets supporting teachers in making informed decisions in a timely manner based on the provided contexts. Unlike transactional teaming, where the interaction is more direct and short-lived, situational teaming and higher levels of teaming could incorporate multiple models, including conventional AI and GenAI, in completing a whole analysis pipeline towards final outputs delivered to teachers. While conventional AI is mastered in performing task-specific requirements, such as a predictive model to identify students’ emotions from video analysis \cite{savchenko_facial_2023}, GenAI can dynamically operate on a range of multimodal data, hold contextual length in analysing multimodal contextual information and offer situation-specific interpretative information in natural language to teachers. Not limited to a single AI model, these combined AI systems could support teaching tasks at the situational level or beyond. For example, \cite{cohn_multimodal_2025} used Otter.ai to automatically transcribe and diarise students’ conversation, used Multi-Task cascaded ConvolutioNal Networks (MTCNNs) for obtaining face detection from students’ video and a classification model, HSEmotion  \cite{savchenko_facial_2023} for predicting emotions, along with GenAI as GPT-4 for synthesising multimodal data to offer situational feedback to teachers in a timeline format. The affordances of GenAI in making language-based reasoning suggest its flexible capabilities in customising outputs according to situations with relevant narratives. Since the AI agent assists teachers by providing information, teachers play a crucial role in critiquing information, internalising knowledge and acting upon it in educational settings. They must possess skills and competencies in designing and operating pedagogical interventions on their own.

\textbf{Operational teaming} (62.5\%) suggests the division of labour between AI and teachers in the joint operation of educational tasks. In traditional AI-powered authoring tools, a fixed set of widgets representing its design constraints, tends to be offered for teachers to initiate or manipulate. For example, an ITS system offers a button for teachers to assign a student pairing strategy \cite{yang_pair-up_2023}. GenAI-powered systems, on the other hand, could offer naturalistic dialogues for teachers to freely interact with to support task division between agents. To illustrate, \cite{li_generative_2025} deployed a GenAI-powered virtual avatar on a web-based virtual reality platform for job training. Teachers can create prompt-based training scenarios for the system to operate for students. These flexibilities stemming from the stochastic nature of GenAI models pose concerns for teachers over their competencies to operate GenAI according to their pedagogical goals. Similar to the situational level, teachers should cultivate necessary skills to operate AI tools pedagogically, along with acquiring emerging skills such as effective prompting strategies.

GenAI capabilities also profoundly support \textbf{praxical teaming} (2.5\%) as GenAI can be adapted to user preferences through iterative interactions. \cite{jin_teachtune_2025} demonstrates an example of teacher-AI teaming at the praxical level by engaging in how to co-design pedagogical conversational agents (PCAs) for students. Through iterative processes, teachers can define, observe, and refine student profiles and instructional flows for PCAs with simulated students. GenAI can then adapt these PCA flows based on updated instructions, dynamically revising student knowledge states and conversational responses to align with teachers’ intentions. While this may not constitute an explicit \emph{learning mechanism} of GenAI models in updating their internal model parameters via retraining, it is an example of conditioning and adapting the final outputs based on soft learning \cite{NEURIPS2023_73950f0e}. GenAI’s language-based reasoning capabilities enable teachers to engage in natural dialogues that foster deeper cognitive development in everyday practice, for instance, when conceptualising ideas, co-creating pedagogical plans, co-designing learning interactions, or co-producing feedback. Recent developments in agentic AI, particularly multi-agent system implementations in education (e.g. \cite{Yu2026}), also provide significant opportunities for models to update their parameters and capabilities based on immediate experiences. However, their use in practice and research is just emerging without any current strong evidence of impact.

Research has shown evidence supporting complementarity through praxical teaming, covering impacts on both teachers and learners. To illustrate this point, \cite{liu_fast_2024} investigated praxical teaming of teachers and AI in grading students’ assignments. The quantitative results suggest that the praxical approach can decrease grading time by 44\% and increase grading accuracy by 6\% in comparison to manual grading. Teachers also qualitatively reported more enjoyable, easier and faster operations when they had AI-assisted in the processes, attributing to its capabilities in supporting lower cognitive tasks, enabling them to concentrate on higher-order pedagogical tasks. \cite{reza_prompthive_2024}’s study shows aligned results of AI praxically support teachers in producing feedback for students. The co-produced feedback reveals comparable quality to manual feedback and statistically significant learning gains, yet with a significant reduction of feedback production time and perceived workload. While the papers highlight the positive impacts of teachers’ complementary integration of GenAI into practice, challenges remain, including the additional effort required to verify AI outputs and the inherent stochasticity of GenAI models. For instance, as GenAI relies heavily on training corpora, its performance on unseen cases could be deteriorated or unexpected, requiring teacher monitoring, which may in turn increase their workload \cite{liu_fast_2024}. \cite{reza_prompthive_2024} reported AI randomness in producing unpredictable results and participants’ complaints in reengineering prompts to shape GenAI in producing pedagogically aligned results. In addition, these studies do not necessarily report on the impact of such teacher-AI interaction on teacher competence and highlight that the observed results also significantly depend upon the specific competencies of teachers. Altogether, findings suggest an emerging need in supporting teachers to enhance their competencies related to teaming with AI.

Our review reveals a notable absence of empirical evidence on teacher-AI synergistic teaming. As a result, improvements in teacher competence through interaction with AI (compared to teachers or AI systems functioning alone) are under reported. The next section addresses this gap by analysing possible explanations for the limited occurrence of synergy and by drawing on related research to outline conditions that may enable its development.

\section{Towards Teacher-AI Synergy}
The proposed five levels of teaming offer a theoretically grounded framework to conceptualise the degrees of teacher agency, teacher-AI interactions and levels of their adaptation through the lens of AI models’ interaction affordances. The framework can posit a structural roadmap one can use to systematically enhance any system to build up greater cognitive support capabilities within the human-AI system. Any teacher-AI interaction in achieving teaching tasks in educational contexts can be taken and, firstly, positioned in the framework to understand its current state, automatically indicating potential directions of improvement. As discussed earlier, systems designed for different teaching tasks and varying levels of complexity may progress through the human-AI teaming levels in distinct ways. The proposed framework, however, serves as a systematic guide for steering such evolution, avoiding arbitrary adjustments and ultimately supporting the development of synergistic teaming capacities.

While previous research leveraged GenAI across the prior four levels, we find that there is a lack of examples to showcase a synergistic teaming between teachers and AI to support teaching tasks in our reviewed papers. \textbf{Synergistic teaming} refers to the situation where teachers and AI act proactively in mutual and engaging negotiation with equal levels of autonomy. The ability to equate levels of autonomy can give rise to constructive debate, leading to creative solutions. It is not yet clear whether GenAI currently poses capabilities for genuinely \emph{pushing back} against teachers on their instructions or decisions given the sycophantic nature of current state-of-the-field AI models. One line of argument concerns the philosophical stance on whether AI has agency in the first place \cite{floridi_ai_2025} and another is on its current technical capabilities, relying on pre-trained data, fine-tuning approaches to force output generation and optimise user satisfaction, as well as methods for training models without any built-in intentionality. 

As GenAI models are pre-trained on massive data and rely on probabilistic language models, they process prompts through iterative processes: tokenisation (converting inputs into tokens), embedding conversion (mapping tokens into a multidimensional vector space), self-attention analysis (identifying semantic relationships among tokens) and token generation (predicting next probable words), to formulate sentence outputs. These semantic relationship mining and probabilistic models might limit their capabilities to authentically \emph{push back}, as there is no internal epistemic stance of GenAI models, and there is not enough evidence that such a stance might emerge as a product of the training phase. In addition, a critical phase in training Foundational GenAI models is post-training \cite{tie2025survey} tuning, where models are instruction-tuned with human preferences. Different methods like Reinforcement Learning with Human Feedback (RLHF) and Direct Preference Optimisation (DPO) are post-training phases where models are aligned with human preferences. While these have benefits for human alignment, it is arguable that they also lead to emergent features like sycophancy which causes the deterioration of models' ability to disagree and \emph{push back} when engaging with human users to provide more opportunities for human competence development and learning.

However, it might be able to demonstrate its disagreement behaviours if it is explicitly requested for counterarguments and if the inquired counterarguments dominate pre-trained corpora. Alternatively, we hypothesise that GenAI, in combination with conventional AI models or techniques, could be integrated to achieve simulations of co-reasoning and co-adaptation that can allow synergistic teaming between teachers and AI. GenAI offers flexibility in coping with contextual differences, whereas the other models or techniques provide a theoretical ground or epistemic stance for pushing back. Through iterative debateful exchanges and boundary extensions of one another, teachers’ professional development is cultivated, as well as an improvement of GenAI models are provided, reflecting the co-adaptation of both agents. 

In addition, the capacity to push back can be double-edged in that models capable of resisting human input may also disregard valid feedback, making independent decisions without transparency or accountability. In the absence of clear governance mechanisms or ethical oversight, such autonomy could introduce serious risks in educational contexts. These concerns may partly explain why current fine-tuning practices remain conservative, prioritising controllability and alignment over the development of genuinely dialogic or dissent-capable AI systems.

While no synergistic teaming was found in our reviewed papers in teacher-facing AI in Education literature, \cite{vaccaro_when_2024} presents synergistic performance evidence as synthesised from a meta-analysis of over a hundred experimental studies beyond educational fields, highlighting several factors that condition synergy. Results strikingly reveal that in 58\% of the reviewed studies, the performance obtained from human-AI integration was lower than the better performer alone (either human or AI). This offers a counterintuitive argument that synergic interactions are not naturally achieved. One factor influencing synergy could be the \emph{instructional design} shaping human-AI interactions or teaming levels. As only 3\% of the reviewed papers in \cite{vaccaro_when_2024}'s study reported experimenting on subtask delegation of both agents. These low-structured interactions could potentially contribute to low reported synergy gains, emphasising the necessity to systematically design instructions to effectively scaffold human-AI interactions. While we acknowledge that the five levels of teaming do not necessarily represent superior educational outcomes, higher degrees of shared control from both teachers and AI, particularly at the praxical and synergistic levels, reflect deliberate efforts to design richer opportunities for dynamic interactions and foster co-adaptation towards synergy. 

Another factor concerns \emph{contextual factors} such as the nature of the task, which influences the extent to which synergy is likely to occur. In the same study \cite{vaccaro_when_2024}, synergy is more anticipated in simpler content generation tasks, whereas in complex and abstract decision-making tasks, synergy is hardly achieved. This raises a key consideration regarding the definition of success or how performance was measured in each context. It is therefore essential to establish robust and equitable metrics and systematic protocols in evaluating human-AI performance, as the validity of impact assessments heavily relies on the quality and fairness of those metrics and the conducting approaches. However, \cite{vaccaro_when_2024}’s review doesn’t report on the impact of different human-AI teaming studies on human or AI competence before and after their interactions with each other, which makes its value for educational and learning sciences research limited.

Inevitably, \emph{human factors} play a significant role in achieving synergy. The study also shows that synergy is more likely to be accomplished if and only if humans exceed AI performance, not vice versa. This unravels synergistic conditions where human competencies are essential in having higher control and agency in critically judging the extent to which AI results are trustworthy, signifying the need to improve AI literacy of teachers (see UNESCO AI competency framework for teachers by \cite{miao_ai_2024}). Professional development is thereby necessary to nurture teachers’ competence to be AI literate in making sense and operating on information pedagogically and engaging with AI in critical and reflective manners, as well as develop motivation to engage with AI meaningfully, aligning with their own professional goals. 

Furthermore, \emph{GenAI technical capabilities and designs} play a vital role in nurturing synergy as previously elaborated. Especially in the educational context, the development of pedagogically aligned GenAI systems is significant in contributing to task-specific synergy. Relying solely on proprietary and general-purpose GenAI tools may limit the alignment between teachers’ pedagogical goals and GenAI capabilities. These can be achieved through technological advancements of GenAI systems in conjunction with human-centred design development \cite{buckingham_shum_human-centred_2019}, incorporating key stakeholders such as teachers in designing interaction scaffolds, promoting synergistic interactions.

Collectively, these discussions remain at an early stage, both in the development of GenAI and in its pedagogical and ethical integration into educational practice, highlighting the need for future empirical investigation.

At this early stage of research, to support institutions seeking to translate this framework into practice, implementation should be approached as a staged, capacity-building process rather than a merely technical deployment challenge. A practical starting point is for institutions to audit existing teacher-AI interactions within their platforms and workflows and map them onto the five levels of teaming to identify their current baseline. This mapping can then guide intentional design decisions, for instance, upgrading transactional tools towards situational or operational teaming by integrating classroom-relevant data streams or providing teachers with goal-setting interfaces. Institutions should also establish structured professional development aligned with AI literacy frameworks such as the UNESCO AI CFT \cite{miao_ai_2024}, ensuring that teachers develop the metacognitive, critical and pedagogical skills necessary for praxical and eventually synergistic interactions. Importantly, effective implementation requires human-centred design processes in which teachers participate in co-designing models, prompts, interaction protocols and decision-support flows to ensure pedagogical alignment and local relevance. Finally, institutions should implement robust evaluation cycles using process analytics, teacher reflections and educationally meaningful outcome measures to iteratively refine the configurations of teacher-AI teaming. 

\section{Conclusion}

Despite growing concerns that GenAI may automate educational tasks, reduce teachers’ opportunities for cognitive engagement and pedagogical practice, and even displace teachers from classrooms \cite{cukurova_promoting_2025}, it also reveals complementary benefits for enhancing teaching practice \cite{roy_chatgpt_2024}. It not only helps achieve tasks productively but also helps articulate how one could effectively complete pedagogical tasks while maintaining agency of both agents towards mutual development. Beyond discussing GenAI’s technical affordances, this chapter introduces a conceptual framework describing five levels of teacher-AI teaming situated within concrete interaction spaces between teachers and AI, particularly GenAI. We present findings from a recent systematic literature review to exemplify the levels and discuss their prominence in the field. The framework provides a lens to interpret diverse use cases and their implications for teachers’ competence. Ultimately, understanding these levels can guide the responsible evolution of GenAI in education; one that augments, rather than replaces, teachers’ professional judgment and fosters genuinely synergistic human-AI teaming.

\begin{acknowledgement}
This research has been funded by the European Union TAICo (‘Teacher-AI Complementarity’) project, financed by the European Commission under HORIZON- CL2-2024-TRANSFORMATIONS- 01-11, project number 101177268.
\end{acknowledgement}
\input{references}

\end{document}

%% file: references.tex
\bibliographystyle{plain}
\bibliography{RefChapter}